\documentstyle[hep93,psfig]{article}
\global\arraycolsep=2pt
\renewcommand{\thefootnote}{\fnsymbol{footnote}}
\begin{document}
\title{\hfill OKHEP-93-08\\
FINITE-ELEMENT QUANTUM ELECTRODYNAMICS}
\author{DEAN MILLER, KIMBALL A. MILTON, and STEPHAN SIEGEMUND-BROKA\\
        \it Department of Physics and Astronomy, The University of
Oklahoma\\
        \it Norman, OK 73019 USA
        \
        }
\abstract{\rightskip=1.5pc
          \leftskip=1.5pc
We apply the finite-element lattice equations of motion for quantum
electrodynamics  to an  examination of
anomalies in the current operators.  By taking explicit lattice
divergences of the vector and axial-vector currents we compute
the vector and axial-vector anomalies in two and four dimensions.
We examine anomalous commutators of the currents to compute divergent
and finite Schwinger terms.  And, using free lattice propagators,
we compute the vacuum polarization in two dimensions and hence
the anomaly in the Schwinger model.}

\maketitle
This paper summarizes the status of the finite-ele\-ment approach to
\footnotetext{To appear in the Proceedings of the International
Europhysics
Conference on High Energy Physics, Marseille, July 22-28, 1993.}
gauge theories, in which the Heisenberg operator equations of motion
are converted to operator difference equations consistent with
unitarity.
A review of the entire program, from quantum mechanics to quantum
field theory, is given in Ref.~\cite{review}.
\vskip 4.0mm
{ \bf \noindent Lattice Propagators}
\vskip 2.0mm
We begin by reminding the reader of the form of the free
finite-element
lattice Dirac equation:
\begin{eqnarray}
{i\gamma^0\over h}(\psi_{\overline{{\bf
m}},n+1}&-&\psi_{\overline{{\bf
m}},n})
+{i\gamma^j\over\Delta}(\psi_{m_j+1,\overline{{\bf
m}}_\perp,\overline{n}}\nonumber\\
&-&\psi_{m_j,\overline{{\bf
m}}_\perp,\overline{n}})
-\mu\psi_{\overline{{\bf m}},
\overline{n}}=0.
\end{eqnarray}
Here $\mu$ is the electron mass, $h$ is the temporal lattice spacing,
$\Delta$ is the spatial lattice spacing, $\bf m$ represents a spatial
lattice coordinate, $n$ a temporal coordinate,
 and overbars signify forward averaging:
\begin{equation}
x_{\overline m}={1\over2}(x_{m+1}+x_m).
\end{equation}
It is  a straightforward exercise to show that, apart from a
contact term, the free electron Green's function may be expressed as
\begin{eqnarray}
&&G_{{\bf m},n;
{\bf m}',n'}={h\over4\pi}\int_{-\pi/h}^{\pi/h}
d\hat\Omega e^{-ih\hat\Omega(n-n')}\nonumber\\
&&\quad\times{1\over L^3}\sum_{\bf p}e^{i{\bf  p\cdot(m-m')}2\pi/M}
\nonumber\\
&&\quad\times
{\gamma^0\sin h\hat\Omega+(\mu-\mbox{\boldmath
$\gamma$}\cdot\tilde{\bf
p})
h\cos^2 h\hat\Omega/2\over\cos h(\Omega-i\epsilon)-\cos
h\hat\Omega}.
\label{diracgreen}
\end{eqnarray}
Here the lattice momentum is given by
\begin{equation}
\tilde{\bf p}={2{\bf t}\over\Delta},
\quad \tilde\omega=\tilde p^0=\sqrt{\tilde {\bf p}^2+\mu^2},\quad
({{\bf  t_p}})_i=\tan p_i\pi/M.
\label{tp}
\end{equation}
The quantity  $\Omega$ is related to $\tilde\omega$ by
\begin{equation}
\tilde\omega={2\over h}\tan{h\Omega\over2}.
\label{omega}
\end{equation}
We take $M$, the number of lattice points in a given spatial
direction, to be odd, so that $\psi$ is periodic on the spatial
lattice.

A similar expression can be derived for the free photon propagator.

This propagator (\ref{diracgreen}) has been used to perform a
calculation
of the vacuum polarization in two-dimensional QED, with a result
consistent
with the anomaly in the Schwinger model, $e^2/\pi$.
Some typical results are shown in Fig.\ 1.


\begin{figure}
   \centerline{\psfig{figure=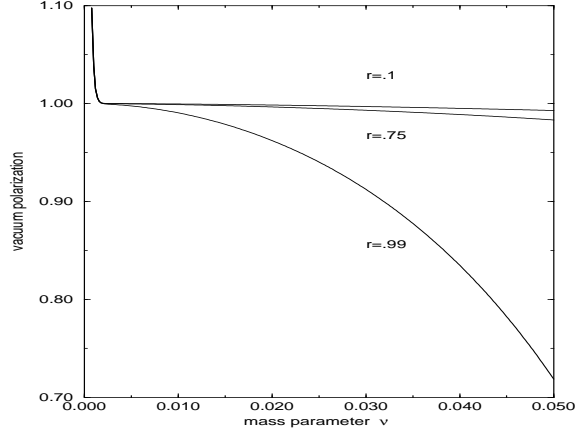,height=5cm,width=6cm}}
   \caption{Plot of the lattice vacuum polarization
$\Pi(0)/(e^2/\pi)$  for $M=2533$ as a function of $\nu
=\mu\Delta/2$.  Shown are curves with $r=h/\Delta=0.1$, 0.75, 0.99.}
\end{figure}

A similar calculation of the anomaly in four-dimen\-sional
electrodynamics
is in progress.

\vskip 4.0mm
{\bf \noindent Interactions}
\vskip 2.0mm

Interactions of an electron with a background electromagnetic field
is given in terms of a transfer matrix $T$:
\begin{equation}
\psi_{n+1}=T_n\psi_n,
\end{equation}
which is to be understood as a matrix equation in $\overline {\bf
m}$.
Explicitly, in the gauge $A^0=0$,
\begin{equation}
T=2U^{-1}-1,\quad
U=1-{ih\mu\gamma^0\over2}+{h\over\Delta}\gamma^0
\mbox{\boldmath$\gamma$}
\cdot\mbox{\boldmath${\cal D}$},
\end{equation}
where
\begin{eqnarray}
{\cal D}^j_{{\bf
m,m'}}&=&(-1)^{m_j+m_j'}[\epsilon_{m_j,m_j'}\cos\hat
\zeta_{m_j,m'_j}\nonumber\\
&+&i\sin\hat\zeta_{m_j,m'_j}]\sec\zeta^{(j)}\,
\delta_{{\bf m_\perp,m'_\perp}}.
\label{eq:d}
\end{eqnarray}
Here
\begin{equation}
\epsilon_{m,m'}=\left\{\begin{array}{ll}1,&m'>m,\\
 0,&m'=m,\\
-1,&m'<m,\end{array}\right.
\end{equation}
and (the following are local and unaveraged in ${\bf m}_\perp$, $n$)
\begin{equation}
\zeta_{m_j}={e\Delta\over2}A^j_{\overline{m_j-1}},\quad \zeta^{(j)}=
\sum_{m_j=1}^M\zeta_{m_j},
\end{equation}
and
\begin{equation}
\hat\zeta_{m_j,m_j'}=\sum_{m_j''=1}^M{\rm sgn}\,(m_j''-m_j){\rm
sgn}\,(m_j''-m_j')\zeta_{m_j''},
\end{equation}
with
\begin{equation}
{\rm sgn}\,(m-m')=\epsilon_{m,m'}-\delta_{m,m'}.
\end{equation}
Because ${\cal D}$ is anti-Hermitian, it follows that $T$ is unitary,
that
is,
that $\phi_{{\bf m},n}=\psi_{\overline{\bf m},n}$ is the canonical
field variable satisfying
the canonical anticommutation relations.

It is instructive, and very simple, to consider the Schwinger model,
that is the case with dimension $d=2$ and mass $\mu=0$.  We set
$h=\Delta$
because the light-cone
aligns with the lattice in that case.  Then we
see that the transfer matrix for positive or negative chirality,
that is, eigenvalue of $i\gamma_5=\gamma^0\gamma^1$ equal to $\pm1$,
is
\begin{equation}
T_\pm={1\pm{\cal D}\over1\mp{\cal D}}.
\end{equation}
It is an immediate consequence of (\ref{eq:d}) that
\begin{eqnarray}
(T_+)_{m,m'}&=&\delta_{m,m'+1}e^{2i\zeta_{m}},\label{plus}\\
(T_-)_{m,m'}&=&\delta_{m+1,m'}e^{-2i\zeta_{m'}},\label{minus}
\end{eqnarray}
which simply says that the $+$ ($-$) chirality fermions move on the
light-cone
to the right (left), acquiring a phase proportional to the vector
potential.

Using (\ref{plus}) and (\ref{minus}) the anomaly in the Schwinger
model is
easily computed.  The corresponding calculation in four-dimensional
electrodynamics will be presented elsewhere.

\bigskip\bigskip\vfill\eject
{\bf \noindent Anomalous Current Commutators}
\vskip 1.0mm

It is extremely interesting to compute commutators of the
gauge-invariant
lattice current, and compare with the anomalous commutators in the
continuum:
\begin{equation}
\langle[J^0(0,{\bf x}),{\bf J}(0)]\rangle=iS\mbox{\boldmath$
\nabla$}\delta
({\bf x})
+{id}\mbox{\boldmath$ \nabla$}\nabla^2\delta({\bf x}),
\end{equation}
where $S$ is the quadratically divergent Schwinger term
\cite{Schwinger}, and $d=1/12\pi^2$
for the
Bjorken-Johnson-Low regularization \cite{BJL}.
We have carried out a straightfoward evaluation of the
current-current commutators on the
lattice, and have performed a fit with lattice delta functions.
The results of  such fits for the
coefficients
of the first three odd derivatives of delta functions
are shown in Fig.\ 2 for various lattice sizes.
The coefficient of the first derivative is roughly consistent with
the
Schwinger
result \cite{Schwinger}. Similarly, the result for the
coefficient of the third derivative term in the commutator is in
decent
agreement with the BJL  result \cite{BJL};
however, in both cases, there seem to be significant discrepancies.


\begin{figure}
   \centerline{\psfig{figure=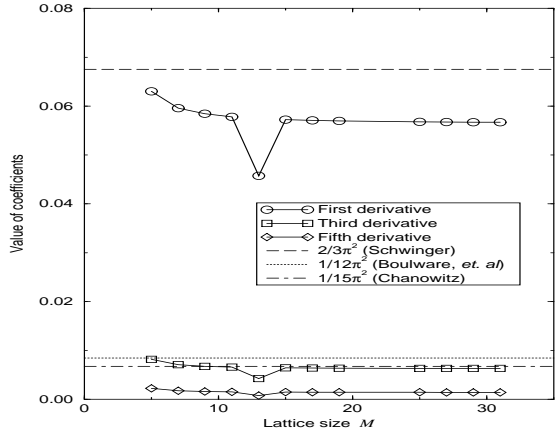,height=5cm,width=6cm}}
   \caption{Coefficients of three spectral fit for different
lattice sizes.
($\mu=0$) For the first derivative term, the coefficient shown
is $S\Delta^2$.}
\end{figure}


\vskip 1.0mm
{\bf \noindent Acknowledgements}
\vskip 1.0mm
This work was supported in part by the U.S. Department of Energy and
the
U.S. Department of Education.

\end{document}